\documentclass{PoS}

\title{Preparing old and recent radio source tables for the VO age: Current status}

\ShortTitle{Preparing old and recent radio source tables for the VO age: Current status}

\author{\speaker{Heinz Andernach}~$^{a, b}$ \\
  \llap{$^a$}Argelander-Institut f\"ur Astronomie (AIfA), Universit\"at Bonn, Auf dem H\"ugel 71\\
D-53121 Bonn, Germany \\
\llap{$^b$}On leave of absence from Depto.\ de Astronom\'{\i}a, 
Universidad de Guanajuato, AP 144\\
Guanajuato, CP 36000, Gto, Mexico\\
E-mail: \email{heinz@astro.uni-bonn.de}}

\abstract{
Independent of established data centers, and partly for my own research,
I have been collecting the tabular data from nearly 1500 articles concerned
with radio sources. Optical character recognition (OCR) was used to
recover tables from nearly 600 of these. Tables from only 44 percent 
of these articles are available in the CDS or CATS catalog collections. 
This fraction is 62 percent for articles with over 100 sources.  
Surprisingly, these fractions are not better for articles published
electronically since 2001, perhaps partly due to the fact that
often tabular data are published in formats not useful for direct machine reading.
The databases Simbad and NED recognize only about 60 percent
of the bibliographic references corresponding to the existing electronic
radio source lists, and the number of objects associated with these
references is much smaller still.  Both, object databases like NED and Simbad,
as well as catalog browsers (VizieR, CATS) need to be consulted to
obtain the most complete information on radio sources. More human
resources at the data centers and better collaboration between authors,
referees, editors, publishers, and data centers are required to improve
the flow of tabular data from journals to public databases. 
Current efforts within the Virtual
Observatory (VO) project, to provide retrieval and analysis tools for
different types of published and archival data stored at various sites,
should be balanced by an equal effort to recover and include large
amounts of published data not currently available in this way.
If human resources can be found, the data sets collected by the author 
will be made available for the preparation of metadata necessary for their 
ingression into catalog browsers.
}

\FullConference{Panoramic Radio Astronomy: Wide-field 1-2 GHz research on galaxy evolution\\
                 June 2-5 2009\\
                 Groningen, the Netherlands}

\begin{document}

\section{Introduction}

Since 1989, motivated by a lack of data on radio sources in
NED (nedwww.ipac.caltech.edu), Simbad (simbad.u-strasbg.fr), and the CDS 
catalogs (cdsarc.u-strasbg.fr/viz-bin/ftp-index), I collected electronic
tables of radio sources and/or extragalactic objects that (a) contained 
$\gtrsim$50 records, and (b) were unavailable from data centers at the time. 
Many of these catalogs were later incorporated in the CDS archive and the VizieR
catalog browser (vizier.u-strasbg.fr, \cite{OCHS00}), as well as in the 
CATS catalog browser (cats.sao.ru, \cite{VER09}).
My collection of data tables from currently $\sim$2950 articles is
described in more detail (as of Sep.~2008) in \cite{AND09}. Here
I concentrate on the subset of tables from 1494 articles dealing
with radio sources. Tables from about 580 of these, published 1950--1999
and with a total of $\sim$200,000 records, were recovered with optical
character recognition (OCR) software. Since 2004 the collection has grown
by $\sim$110 articles/yr, half of these published before 1999 and recovered 
via OCR, sometimes correcting the raw OCR results of the Astrophysics Data System
(adsabs.harvard.edu/cgi-bin/signup\_ocr). Tables from a further
$\sim$110 articles at CDS/VizieR were not duplicated into my
collection. Thus, the total number of articles with existing electronic 
radio source tables is currently $\sim$1600.
The cumulative size distribution of the latter is shown by the continuous
line in Figure~1. Here, {\it size} is typically the number of objects in a
catalog, but may also be the number of flux density measurements,
e.g.\ in case of various observing frequencies or epochs.  The size distribution 
closely follows a power law with a slope of $-$0.70.  Such power laws are 
known in biometrics as Zipf's or Lotka's laws.
For sizes $\lesssim$100 records a ``collection bias'' sets
in, caused by working from the bigger towards the smaller lists.

\section{Comparison with other catalog browsers}

The largest collection of astronomical catalogs is maintained by CDS 
at Strasbourg, where, as of Sept.~2009, $\sim$8000 catalogs are accessible 
for download, and 7560 are also available for cone searches through the VizieR
browser. The dashed line in Fig.~1 indicates the
size distribution of radio catalogs in the CDS collection, which is
complete for radio catalogs of $\gtrsim$10,000 records. The dotted
line shows the same for CATS, which is complete for $\gtrsim$5000 records.

The left side of Table~1 lists the number of items in each collection of 
radio catalogs, followed by the percentage in number of catalogs, their
total number of records, and their percentage in number of records,
as of 1-Sep-2009. HA stands for the author, and the third row refers
to catalogs which are in the author's, but not in the CDS or CATS
collections. I am not aware of any electronic radio catalog outside the
HA+CDS collection. For those electronic radio source catalogs with more 
than 100 records,
only 52\,\% are in the CDS archive, and only 34\,\% are included in CATS.
While CATS offers more data (records), VizieR offers a larger variety of catalogs, 
i.e.\ more of the smaller source lists.  One needs to search in both services
to obtain the most complete results, and then weed out the duplications
from the 244 catalogs in common between CATS and VizieR. 
More than half of the items I collected are neither
in VizieR nor in CATS. However, the preparation of metadata is 
required for their inclusion in catalog browsers.
Some of these catalogs require a further effort (e.g.\ insert missing
absolute coordinates from their object names), and a still
smaller fraction of these catalogs, e.g.\ literature compilations or those 
containing only derived parameters, may be unsuitable for inclusion into 
catalog browsers.

\begin{figure}[t!]
\vspace*{-3mm}
\begin{center}
\includegraphics[width=0.6\textwidth,angle=-90,viewport=5 5 490 514,clip]{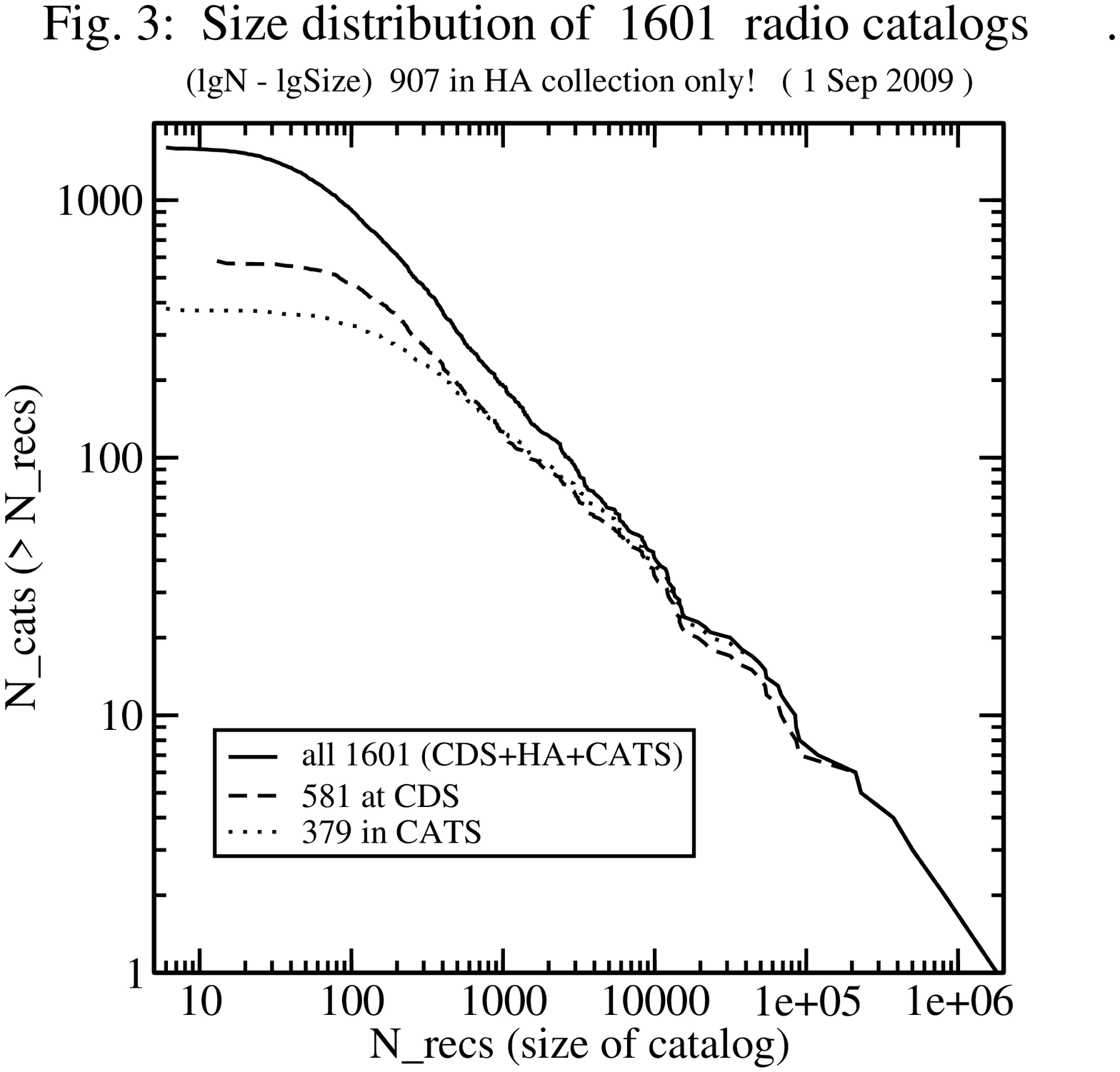}
\caption{Cumulative size distribution of 1601 radio catalogs. For any
given size (in records) listed on the abscissa, the ordinate gives the
total number of catalogs up to that size (as of 1--Sep--2009). Continuous
line: author's collection plus those at CDS and CATS; dashed line:
subset of radio catalogs that are available at CDS; dotted line: subset
of radio catalogs that are available in CATS.}
\label{fig1}
\end{center}
\end{figure}

\begin{table}
\begin{center}
\footnotesize
\label{tab:id}
\tabcolsep2mm
\begin{tabular}{lrrrr|rrrr}
\hline
    & \multicolumn{4}{c}{All radio source tables}   & \multicolumn{4}{c}{Tables with $>$50 records publ. since 2001} \\
\hline
Collection        &     Ncats  & \%cats & Nrecords  & \%recs   &   ~~~Ncats  & \%cats & Nrecords  & \%recs \\
\hline
VizieR (not INPREP)  &    581  &  36.3 & 5,269,531 &  90.6 &  237 &  65.7 & 1,634,877 &   94.1 \\
CATS collection      &    379  &  23.7 & 5,449,164 &  95.0 &   57 &  15.7 & 1,530,200 &   88.1 \\
HA collection ONLY   &    901  &  56.3 &   209,332 &   3.6 &  103 &  28.5 &    81,000 &    4.7 \\
HA+CDS collections   &   1601  & 100.0 & 5,736,178 & 100.0 &  361 & 100.0 & 1,737,474 &  100.0 \\
\hline
\end{tabular}
\caption{Left part: Presence of all electronic radio source catalogs in various 
collections; right part: same for the subset of tables with $>$50 records and 
published since 2001.}
\end{center}
\end{table}

\section{Recent Catalogs published in Electronic Journals}

One might think that the vast majority of the catalogs missing from
the public catalog browsers is outdated, and that in recent years
there is a continuous ``flow'' of electronically published source tables
into the data centers and catalog browsers.  However, the right-hand half 
of Table~1 shows that this is not the case. Of the 361 radio source tables with 
$>$50 records, and published since 2001, mostly in electronic journals, 
only $\sim$66\% (albeit with 94\% of the total number of records) are in 
VizieR. This may be due to the fact that journal tables are often published
only in LaTeX, PS or PDF format, or they are not readily machine-readable 
for various reasons, e.g.\ lack of metadata, columns not 
aligned, or a mixture of ASCII, HTML and LaTeX codes (cf.\ \cite{AND09}).
Despite the fact that the 32\% of tables missing from VizieR and CATS ``only'' 
contain 5\% of all records, these offer flux measurements at a larger variety 
of frequencies and epochs than the big catalogs.

\begin{table}
\begin{center}
\footnotesize
\label{tab2}
\begin{tabular}{lccc|ccc}
\hline
    & \multicolumn{3}{c}{~~~All 1565 radio source tables~~~~~}   & \multicolumn{3}{c}{354 tables $>$50 records, publ. since 2001} \\
\hline
Service~~~~~~~~~~  &   Ncats  & ~~~~\%cats & \%recs & ~~~~Ncats  & ~~~~~~\%cats & \%recs \\
\hline
Simbad   &    874  &  55.8 &  4.0 &       ~~~~209 &  59.0 &  4.8 \\
NED      &    984  &  62.9 & 20.2 &       ~~~~194 &  54.8 & 46.0 \\
\hline
\end{tabular}
\caption{Simbad and NED's coverage for 1565 articles with radio source data (left part)
and 354 articles with $>$50 records published since 2001. Columns are: number of 
articles with at least one object in the service, percentage of all articles, 
percentage of total number of objects in the service compared to the total number of 
records in the catalogs.}
\end{center}
\end{table}

\section{Coverage of Tabular Data in Databases}

For most astronomers the major source of information on objects are databases 
like Simbad or NED, whose content are {\it different} from the catalog collections
and browsers described above (cf.\ \cite{AND09}).  Thus I tried
to estimate the fraction of tabular data covered by these databases.
For 1565 of the radio source tables (exluding theses and unpublished items)
both NED and Simbad were interrogated for the number of objects linked
to their 19-digit {\it refcode}, and compared with the full catalog size. 
This is not always reasonable, as some tables give several records
per astronomical objects, while other tables may not be of relevance
(e.g.\ Galactic objects  in NED). Moreover, for some of the largest
datasets (e.g.\ NVSS, PKSCAT90) a NED search {\it by refcode} does not work.
The results are listed in Table~2 for two different samples of references.
Despite the caveats, a severe underrepresentation of published data is obvious: 
regardless of size or publication date, no more than $\sim$60\% of the {\it refcodes} 
are recognized, and the fraction of objects included is still smaller.   \\[-3ex]

\section{Conclusions}

A vast amount of the simplest form of {\it published} data (e.g.\ flux measurements)
now exists in electronic form, but is unavailable in public web services. 
About half of these are readily accessible from electronic journals, and
another half was painfully collected or recovered via OCR by the author.
The preparation of these source lists for the ingression in catalog
browsers is beyond the possibilities of the author and requires more
human resources. A list of available tables can be viewed at 
www.astro.ugto.mx/$\sim$heinz/cats.sum, and individual data sets are
available on request. \\[-3.5ex]

\acknowledgments
I thank the many students and secretaries who helped to recover journal tables 
via OCR and proofread them. Recently I enjoyed
the hospitality at AIfA Bonn, Germany, with support from Mexican CONACyT 
grant 81356, and from German DFG through grants RE1462/2 and TRR33. \\[-3.5ex]

\end{document}